\documentstyle[prl,aps,preprint]{revtex}

\begin{document}
\title
{Reply to Comment by Dhar}
\maketitle

In our paper ``Can disorder induce a finite thermal conductivity in 1D lattices?" (Phys. Rev. Lett. {\bf 86}, 63-66 (2001))\cite{LZH01}, we have pointed out two important things. First, the disorder can induce a finite thermal conductivity in lower temperature regime. Second, our numerical calculations with Noo\'se-Hoover thermostats shown that a unique nonequilibrium stationary state may not exist in a disorder harmonic chain.

In his comment, Dhar\cite{Dhar} claims that the second point is not true. He mentioned that the existence and uniqueness of a nonequilibrium stationary state has been proved by Lebowitz {\it et al} \cite{Lebowitz71}. We should stress here that, our observation does not contradict  the proof of Lebowitz {\it et al}.  
In Ref.\cite{Lebowitz71}, the authors only proved that when such a (mass) disordered 
harmonic chain is placed in contact with stochastic reservoirs of the Langevin type, a nonequilibrium stationary state can be reached. However, they didn't prove the existence and uniqueness of the nonequilibrium stationary state for the general case, namely, when the chain is in contact with general thermostated reservoirs. As pointed out recently by Bonetto, Lebowitz and Rey-Bellet\cite{Lebowitz00} that ``{\it for general thermostated reservoirs the problem seems to be mathematically out of reach at the present time}". So, the existence and uniqueness of a stationary state in a disordered harmonic chain for a general thermostated reservoir is still an open question. 
Most recently, Parisi\cite{Parisi} shows that,  the time needed to equilibrate in a harmonic chain is infinite. Our numerical results agree with this conclusion, viz., at any finite time, one cannot obtain a definite steady state starting from different initial conditions.

Dhar's claim that our results might be caused by the insufficient equilibration time also seems to be unlikely. In Fig. 4(a)\cite{LZH01}, we have shown that the temperature profile at two different time scales $t=10^6$ and $t=10^7$ are almost identical (see also the inset of this figure) when starting from the same initial condition.

\bigskip
\bigskip

\begin{flushleft}
B Li$^{1*}$, H Zhao$^2$, and B Hu$^{3,4}$\\
$^1$ Department of Physics, National University of Singapore, 119260 Singapore.\\
$^2$ Department of Physics, Lanzhou University,  730000 Lanzhou, PR China\\
$^3$ Department of Physics and Center for Nonlinear Studies, Hong Kong Baptist University, Hong Kong, China\\
$^4$ Department of Physics, University of Houston, Houston, Texas 77204-506.
\\
$^*$Author to whom correspondence should be addressed. Email: phylibw@nus.edu.sg
\end{flushleft}

\begin{thebibliography}{9}
\bibitem{Dhar}
A. Dhar,
Comment on "Can disorder induce a finite thermal conductivity in 1D lattices?". Cond-mat/0106349
\end{thebibliography}
\end{document}